\documentclass[modern,trackchanges,twocolumn]{aastex62}

\usepackage{subfiles}
\usepackage{amsmath,amssymb}
\usepackage{gensymb}
\usepackage{rotating}

\mathchardef\mhyphen="2D 

\shorttitle{SOFIA/FORCAST Galactic Center Survey Overview}
\shortauthors{Hankins et al.}

\begin{document}

\title{SOFIA/FORCAST Galactic Center Legacy Survey: Overview}

\correspondingauthor{Matthew J. Hankins}
\email{mhankins@astro.caltech.edu}

\author[0000-0001-9315-8437]{Matthew  J. Hankins}
\affil{Division of Physics, Mathematics, and Astronomy, California Institute of Technology, Pasadena, CA 91125, USA}

\author{Ryan M. Lau}
\affil{Institute of Space \& Astronautical Science, Japan Aerospace Exploration Agency, 3-1-1 Yoshinodai, Chuo-ku, Sagamihara, Kanagawa 252-5210, Japan}

\author{James T. Radomski}
\affil{SOFIA-USRA, NASA Ames Research Center, MS 232-12, Moffett Field, CA 94035, USA}

\author{Angela S. Cotera}
\affiliation{SETI Institute, 189 Bernardo Ave., Mountain View, CA 94043, USA}

\author[0000-0002-6753-2066]{Mark R. Morris}
\affil{Dept. of Physics and Astronomy, University of California, Los Angeles, CA 90095-1547, USA}

\author[0000-0001-8782-1992]{Elisabeth A. C. Mills}
\affil{Department of Physics and Astronomy, University of Kansas, 1251 Wescoe Hall Dr., Lawrence, KS 66045, USA}

\author{Daniel L. Walker}
\affil{National Astronomical Observatory of Japan, 2-21-1 Osawa, Mitaka, Tokyo, 181-8588, Japan}
\affil{Joint ALMA Observatory, Alonso de C\'{o}rdova 3107, Vitacura, Santiago, Chile}

\author{Ashley T. Barnes}
\affiliation{$^{1}$Argelander-Institut f\"{u}r Astronomie, Universit\"{a}t Bonn, Auf dem H\"{u}gel 71, 53121, Bonn, Germany}

\author[0000-0001-8095-4610]{Janet P. Simpson}
\affiliation{SETI Institute, 189 Bernardo Ave., Mountain View, CA 94043, USA}

\author[0000-0002-3856-8385]{Terry L. Herter}
\affil{Department of Astronomy, Cornell University, Space Sciences Bldg, Ithaca, NY 14853-6801, USA}

\author{Steven N. Longmore}
\affil{Astrophysics Research Institute, Liverpool John Moores University, 146 Brownlow Hill, Liverpool L3 5RF, UK}

\author{John Bally}
\affil{Department of Astrophysical and Planetary Sciences, University of Colorado, 389 UCB, Boulder, CO 80309, USA}

\author{Mansi M. Kasliwal}
\affil{Division of Physics, Mathematics, and Astronomy, California Institute of Technology, Pasadena, CA 91125, USA}

\author{Nadeen B. Sabha}
\affil{Institut für Astro- und Teilchenphysik, Universität Innsbruck, Technikerstr. 25, 6020 Innsbruck, Austria}

\author{Macarena Garc\'{i}a-Marin}
\affil{European Space Agency, 3700 San Martin Drive, Baltimore, MD 21218, USA}

\begin{abstract}

The Galactic Center contains some of the most extreme conditions for star formation in our Galaxy as well as many other phenomena that are unique to this region. Given our relative proximity to the Galactic Center, we are able to study details of physical processes to a level that is simply not yet possible for more distant galaxies, yielding an otherwise inaccessible view of the nuclear region of a galaxy. We recently carried out a targeted imaging survey of mid-infrared bright portions of the Galactic Center at 25 and 37 $\mu$m using the FORCAST instrument on \textit{SOFIA}. This survey was one of the inaugural Legacy Programs from \textit{SOFIA} cycle 7, observing a total area of 403 arcmin$^2$ (2180 pc$^2$), including the Sgr A, B, and C complexes. Here we present an overview of the survey strategy, observations, and data reduction as an accompaniment to the initial public release of the survey data. We discuss interesting regions and features within the data including extended features near the circumnuclear disk, structures in the Arched Filaments and Sickle \ion{H}{2} regions, and signs of embedded star formation in Sgr B2 and Sgr C. We also feature a handful of less well studied mid-infrared sources located between Sgr A and Sgr C that could be sites of relatively isolated star formation activity. Last, we discuss plans for subsequent publications and future data releases from the survey.

\end{abstract}


\vspace{0.6in}

\section{Introduction} \label{sec:intro}

The environment in the Galactic Center (GC) is unlike any other part of our Galaxy. The region contains high molecular gas densities \citep{Guesten1983}, high temperatures \citep{Morris1983,Guesten1985}, and large turbulent motions \citep{Bally1987}, all inside a deep gravitational potential well \citep{Morris1996}. The conditions in the GC mirror those found in the nuclei of luminous infrared galaxies and high-redshift systems near the peak of cosmic star formation history \citep{Kruijssen2013}, but its relative proximity \citep[$d=8.0\pm0.5~$kpc;][]{Reid1993} enables us to study the physical processes there at a level of detail that is simply not possible in more distant systems \citep[see the recent review by][]{Mills2017}. 

Observations of the GC often challenge theoretical models of star formation, which often break down in this complex region. For example, while the GC comprises less than 0.01\% of the total volume of the Galactic disk, its star formation rate (SFR) \citep[$\sim0.1~\mathrm{M_{\odot}~yr^{-1}}$;][]{Immer2012ISOgal,Barnes2017} is a considerable fraction of the total SFR of the Galaxy \citep[$\sim1.2~\mathrm{M_{\odot}~yr^{-1}}$;][]{Lee2012}. However, the global GC SFR is more than an order of magnitude smaller than one would expect based on scaling relations with its dense molecular gas content \citep{Lada2012,Longmore2013}. The inefficiency of the GC in converting dense gas to stars presents a significant quandary for the region with much broader implications. For example, SFR measurements are universally used as a fundamental diagnostic tool for understanding the underlying physics in galaxies and for understanding galaxy evolution.

Even with the aforementioned star formation deficiency, the stellar inventory of the GC is relatively rich with numerous types of massive evolved stars, such as Luminous Blue Variables and Wolf-Rayet stars \citep[e.g.,][]{Figer2009}. The GC is home to three known massive stellar clusters \citep{Lu2018} - the Arches cluster \citep{Cotera1996}, the Quintuplet cluster \citep{Okuda1990,Nagata1990}, and the Central cluster \citep{Krabbe1991,Krabbe1995} in addition to a large number of massive field stars which are spread throughout the region \citep[e.g.,][]{Muno2006MassiveStarsGC,Mauerhan2010,Dong2012}. The origin of these massive field stars is somewhat of a mystery. While several sources may be former cluster members which have been dynamically ejected or removed due to tidal evaporation, the entire population of these sources cannot be accounted for with these mechanisms \citep{Habibi2014}. Instead, some fraction of the field stars likely originates from a more isolated mode of star formation, and there is evidence of ongoing massive star and stellar cluster formation within parts of the GC \citep[e.g.,][ and references therein]{Barnes2019}. There have also been efforts to identifying possible clusters associated with GC field stars which has proven observationally challenging \citep{Steinke2016,Dong2017}, though there has been recent progress in this area using large proper motion studies \citep{Shahzamanian2019}.

Large extinction toward the GC \citep[A$_V\sim$30; ][]{Fritz2011} makes it impossible to observe massive stars and protostars at optical and ultraviolet wavelengths. Instead, studies of the region have relied on observations at other wavelengths to examine the distribution and birth environment of stars. Numerous studies of star formation in the GC have been conducted with mid-infrared observations between $\sim$3.6 -- 24 $\mu$m \citep[e.g.,][]{Ramirez2008,Yusef-Zadeh2010,An2011,Immer2012ISOgal}. In particular, warm dust emission at $\sim24~\mu$m is a valuable probe for identifying young stellar objects (YSOs) and estimating star formation rates \citep[e.g.,][]{Calzetti2007}. However, the most active regions within the inner $\sim$200\,pc of our Galaxy are strongly saturated in the \textit{Spitzer}/MIPS 24 $\mu$m data \citep{Yusef-Zadeh2009}. Earlier observations with the \textit{Midcourse Space Experiment} \citep[MSX;][]{Egan2003} at 21.3 $\mu$m (Band E) provide unsaturated images, but are relatively low spatial resolution ($\sim$20" or $\sim$0.8 pc) when compared with \textit{Spitzer}/MIPS at 24 $\mu$m ($\sim$6" or $\sim$0.2 pc). The MSX data suffer from significant confusion in these complex regions, which presents a considerable hurdle in our understanding of these very active portions within the GC. Furthermore, the lack of high-quality mid-infrared data in these regions represents an essential, missing piece of the rich multi-wavelength picture of the GC that has emerged over the last decade.

In order to create improved mid-infrared maps of the brightest portions of the inner $\sim$200 pc of our Galaxy, we set out to conduct a targeted survey of regions within the GC using \textit{SOFIA}/FORCAST. This survey was selected as one of the inaugural Legacy Programs in \textit{SOFIA} cycle 7. Observations were obtained at 25 and 37 $\mu$m using the Faint Object infraRed CAmera for the SOFIA Telescope (FORCAST), enabling us to create a high-resolution (FWHM$\sim$2.3" or $\sim$0.07 pc at 25 $\mu$m and FWHM$\sim$3.4" or $\sim$0.1 pc at 37 $\mu$m) mosaic of portions within the GC including the Sgr A complex and other prominent star forming regions such as Sgr B and Sgr C.

In this paper we present a description of the observations along with the survey strategy and an initial look at the data set with various regions of interest highlighted. In section \ref{sec:obs}, we provide details on the observations and information regarding the creation of the mosaics presented in this work. In section \ref{sec:Discussion}, we highlight and discuss regions and sources that stand out as particularly interesting. These overviews are not intended to be a full and complete analysis of the objects discussed but rather to emphasize areas where this data set is providing an enhanced view of regions within the GC. We are planning follow-up papers for several of these features, as discussed in the text below. Finally, we provide a summary and future outlook in section \ref{sec:Summary}.

\section{Observations and Data Reduction}\label{sec:obs}

Observations of our mid-infrared GC survey were carried out with the 2.5 m telescope aboard the Stratospheric Observatory for Infrared Astronomy \citep[SOFIA;][]{Young2012} using the FORCAST instrument \citep{Herter2012}. FORCAST is a $256 \times 256$ pixel dual-channel, wide-field mid-infrared camera with a field of view (FOV) of $3.4'\,\times\,3.2'$ and a plate scale of $0.768''$ per pixel. The two channels consist of a short-wavelength camera (SWC) operating at 5 -- 25 $\mu\mathrm{m}$ and a long-wavelength camera (LWC) operating at 28 -- 40 $\mu\mathrm{m}$, and the instrument is capable of observing with both cameras simultaneously using a dichroic beam splitter.

\begin{figure*}[ht!]
\centering
\hspace*{-0.3in}\includegraphics[width=160mm,scale=1.0]{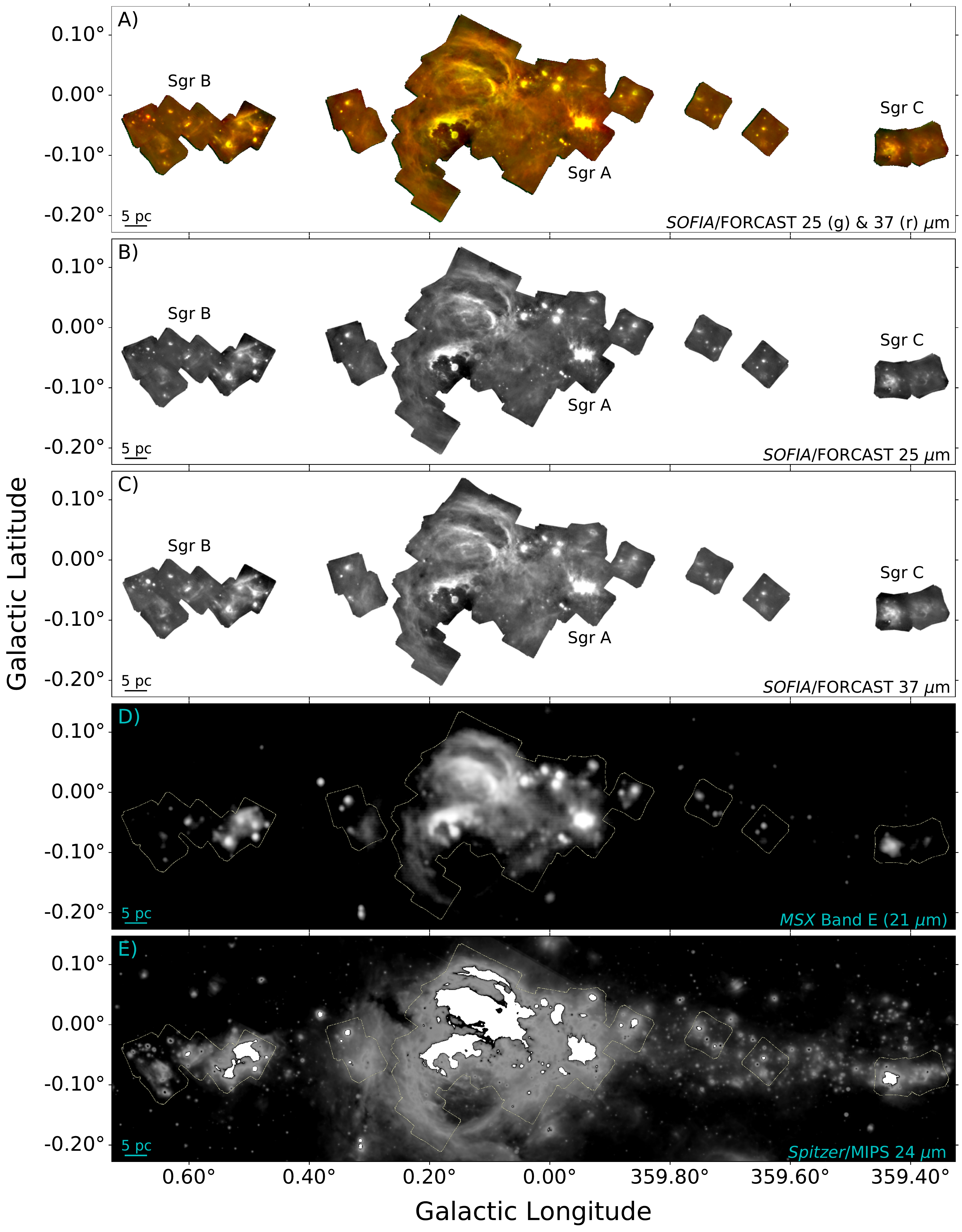}
\caption{{\footnotesize A) The \textit{SOFIA}/FORCAST GC survey mosaic created using the 25.2 (green) \& 37.1 (red) $\mu$m maps. Well known regions including Sgr A, B, and C are labeled here and in panels B \& C. B) The 25.2 $\mu$m \textit{SOFIA}/FORCAST survey mosaic shown in greyscale. C) The 37.1 $\mu$m \textit{SOFIA}/FORCAST survey mosaic shown in greyscale. D) The \textit{MSX} band E (21 $\mu$m) data that were used for planning the FORCAST observations. The stretch of the data shows the approximate depth of the FORCAST observations for comparison and the footprint of the survey area. E) The 24 $\mu$m \textit{Spitzer}/MIPS mosaic of the GC. Several of the brightest infrared features are hard saturated (shown in white). There are additional quality issues with portions of the map neighboring extended hard saturated regions which have high backgrounds and suffer from other bright source effects \citep[see][]{Hinz2009}. The stretch of this figure demonstrates the relative depth of the MIPS mosaic and shows numerous sources outside of the FORCAST survey footprint which have useful data.}}
\label{fig:fig1}
\end{figure*}

\subsection{SOFIA/FORCAST Survey Plan}\label{subsec:obsPlan}

The \textit{SOFIA}/FORCAST Galactic Center survey was designed to map out infrared bright regions within the inner $\sim$200 pc of the Galaxy. We chose to conduct the survey using the FORCAST 25 $\mu$m filter because it is close in wavelength to the MIPS 24 $\mu$m filter and only a slight color-correction ([24]-[25]=0.15 mag for a Vega-like source) is needed to compare photometry between the two data sets where possible. The FORCAST and MIPS data sets are complementary since the higher bright source limit of FORCAST allows for observations of objects and regions where the MIPS data are saturated, while the higher sensitivity of the MIPS probes fainter sources than FORCAST can detect in reasonable integration times. Our nominal sensitivity per field was selected to achieve a 5-$\sigma$ point source depth of 250 mJy at 25 $\mu$m which is equivalent to a 3-$\sigma$ extended source depth of 1200 MJy sr$^{-1}$). This imaging depth allows for detection of class I YSOs (age $<$1 Myr) down to a mass of $\sim$6 M$_{\odot}$ based on the mean 24 $\mu$m flux of YSO models from \cite{Robitaile2006}. This imaging depth is also comfortably below the MIPS hard saturation limit for the existing 24 $\mu$m GC map ($\sim$400 mJy for point sources or $\sim$2300 MJy sr$^{-1}$ for extended emission) allowing for comparison of common sources between the MIPS and FORCAST surveys.

In addition to the 25 $\mu$m filter, we used FORCAST's internal dichroic beam-splitter to enable simultaneous observation with the long wavelength camera which results in only modest loss in the short wavelength channel.\footnote{Throughput of the dichroic for the SWC from 11-25 $\mu$m is 85\%, while the throughput of the LWC from 25-40 $\mu$m is only 40\%} To give the greatest wavelength coverage, we selected the longest wavelength filter available on FORCAST (37.1 $\mu$m) to be paired with the 25 $\mu$m filter observations. 

Based on the integration times determined by the 25 $\mu$m observations, the corresponding 37 $\mu$m 5-$\sigma$ point source depth was 550 mJy and the equivalent 3-$\sigma$ extended source depth was 2200 MJy sr$^{-1}$. The nominal spatial resolution of the 25 and 37 $\mu$m filters in this observing mode are 2.3" and 3.4", respectively.\footnote{https://www.sofia.usra.edu/science/proposing-and-observing/observers-handbook-cycle-7/5-forcast}

Field pointings for the survey were planned using the \textit{MSX} band E data (Figure \ref{fig:fig1}). Additionally, we compared the survey footprint with the aforementioned 24 $\mu$m \textit{Spitzer}/MIPS data in order to optimize coverage of the hard saturated sources. In total our survey plan covered more than 99\% of the hard saturated area in the innermost 200 pc of the Galaxy (Figure \ref{fig:fig1}).

\subsection{SOFIA/FORCAST Cycle 7 Observations}\label{subsec:FORCASTobs7}

A total of 35 fields were observed during SOFIA cycle 7 as part the FORCAST Galactic Center Legacy Program (Program ID: 07-0189; PI: Hankins). Here, we provide a brief description of the observations, with further details provided in Table \ref{tab:tab1}. All 35 fields were observed during the annual SOFIA southern hemisphere deployment to Christchurch, NZ. Observations for the program were spread over 8 total flights which occurred between 1 July 2019 and 11 July 2019. 

Chopping and nodding was used to remove the sky and telescope thermal backgrounds for all observations. This technique requires off fields which are devoid of emission to properly subtract the background emission. The GC is a particularly complex emitting environment, and there were instances where we did not have perfectly blank sky to nod and chop onto within telescope limits. In these cases, we selected the `best' available off field which minimized the number of sources (most frequently an individual source within the expected detection limits). We employed the C2NC2 observing mode to allow use of the full FORCAST FOV, even though this mode comes with considerable overheads due to the necessary three off-source positions. Dithering was used to remove bad pixels and mitigate response variations. We employed different dither patterns for several of the fields in order to minimize observing inefficiencies related to the C2NC2 observing mode, which does not adversely impact the data quality between fields. 

On average we achieved 462s of integration for each field in the 25 $\mu$m filter and an average of 423s for the 37 \micron\ filter. A listing of integration times for each field individually can be found in Table \ref{tab:tab1}. The estimated total photometric errors for the 25 \micron\ and 37 \micron\ data are $\sim$10\%. Two fields (7 \& 11) have integration times which are substantially below the average, and were cut short due to scheduling issues and/or adjustments that were needed in flight. However, the depth of these images is still suitable for some science goals (corresponding to a 250 mJy point source detection of $\sim$3.7-$\sigma$ rather than 5-$\sigma$). In addition to these low-SNR fields, there were three additional fields that were planned as part of the survey but not observed in cycle 7 due to time and observability constraints.

\begin{deluxetable*}{ccccccccc}
\tablecaption{\textit{SOFIA}/FORCAST Observation Details}
\tabletypesize{\scriptsize}
\tablehead{
\colhead{Field ID} & \colhead{Field Center (l,b)} & \colhead{Sky Angle ($\degree$)$^*$} & \colhead{25$\mu$m t$_{\mathrm{int}}$ (s)} & \colhead{37$\mu$m t$_{\mathrm{int}}$ (s)} & \colhead{Chop Angle ($\degree$)$^{\dagger}$} & \colhead{Chop Throw (")} & \colhead{Comment$^{\ddagger}$} & \colhead{AOR ID}
}
\startdata
1 & (359.376, -0.080) & 148.3 & 402 & 369 & 88 & 140 & Affected & 07\_0189\_1 \\
2 & (359.429, -0.087) & 130.1 & 385 & 353 & 52 & 115 &  Affected & 07\_0189\_2 \\
3 & (359.641, -0.062) & 83.5 & 404 & 315 & 60 & 160 &  Post-fix & 07\_0189\_3 \\
5 & (359.737, -0.018) & 99.7 & 461 & 360 & 308 & 165 &  Post-fix & 07\_0189\_5 \\
6 & (0.310, -0.058) & 153.1 & 458 & 415 & 110 & 135 &  Post-fix & 07\_0189\_6 \\
7 & (0.337, -0.024) & 138.9 & 217 & 192 & 230 & 130 &  Post-fix & 07\_0189\_7 \\
8 & (0.584, -0.053) & 174.3 & 433 & 404 & 200 & 190 &  Affected & 07\_0189\_8 \\
9 & (0.631, -0.041) & 178.2 & 614 & 626 & 220 & 160 &  Nominal & 07\_0189\_9 \\
10 & (0.641, -0.092) & 165.1 & 528 & 480 & 77 & 200 &  Affected & 07\_0189\_10 \\
11 & (0.674, -0.051) & 146.2 & 210 & 176 & 130 & 210 &  Affected & 07\_0189\_11 \\
12 & (0.534, -0.072) & 163.3 & 414 & 387 & 80 & 160 &  Affected & 07\_0189\_35 \\
13 & (0.496, -0.051) & 184.7 & 522 & 532 & 60 & 160 &  Nominal & 07\_0189\_36 \\
\hline
A & (359.867, -0.007) & 100.0 & 478 & 400 & 240 & 175 &  Affected & 07\_0189\_12 \\
B & (359.942, 0.027) & 130.2 & 486 & 454 & 270 & 160 &  Affected & 07\_0189\_13 \\
C & (359.931, -0.019) & 85.6 & 440 & 368 & 265 & 210 &  Affected & 07\_0189\_14\\
D & (359.934, -0.067) & 88.3 & 458 & 415 & 5 & 140 &  Affected & 07\_0189\_15 \\
E & (359.970, -0.019) & 143.6 & 393 & 319 & 245 & 210 &  Affected & 07\_0189\_16 \\
F & (359.975, -0.064) & 165.5 & 472 & 433 & 352 & 190 &  Affected & 07\_0189\_17 \\
G & (0.015, -0.022) & 161.0 & 492 & 399 & 250 & 205 &  Affected & 07\_0189\_18 \\
H & (0.014, -0.078) & 188.5 & 527 & 590 & 28 & 200 &  Affected & 07\_0189\_19 \\
I & (0.040, -0.124) & 183.4 & 505 & 458 & 45 & 180 &  Affected & 07\_0189\_20 \\
K & (0.051, -0.008) & 172.3 & 505 & 397 & 260 & 200 &  Affected & 07\_0189\_22 \\
L & (0.056, -0.053) & 183.4 & 524 & 451 & 98 & 210 &  Affected & 07\_0189\_23 \\
M & (0.065, -0.089) & 164.8 & 493 & 447 & 118 & 173 &  Post-fix & 07\_0189\_24 \\
O & (0.101, -0.023) & 181.0 & 472 & 451 & 78 & 210 &  Affected & 07\_0189\_26 \\
P & (0.102, -0.071) & 136.4 & 525 & 511 & 95 & 145 &  Nominal & 07\_0189\_27 \\
Q & (0.145, 0.006) & 188.0 & 518 & 500 & 190 & 210 &  Affected & 07\_0189\_28 \\
R & (0.200, 0.023) & 174.9 & 427 & 387 & 225 & 210 &  Affected & 07\_0189\_29 \\
S & (0.182, -0.018) & 176.6 & 469 & 430 & 180 & 200 &  Affected & 07\_0189\_30 \\
T & (0.223, -0.043) & 159.2 & 492 & 479 & 205 & 200 &  Affected & 07\_0189\_31 \\
U & (0.223, -0.089) & 148.4 & 468 & 437 & 82 & 210 &  Affected & 07\_0189\_32 \\
V & (0.216, -0.134) & 175.4 & 505 & 458 & 75 & 140 &  Post-fix & 07\_0189\_33 \\
W & (0.190, -0.169) & 180.9 & 482 & 500 & 80 & 140 &  Post-fix & 07\_0189\_34 \\
X & (0.142, -0.045) & 181.6 & 484 & 451 & 60 & 180 &  Post-fix & 07\_0189\_37 \\
Y & (0.175, -0.067) & 170.6 & 504 & 471 & 35 & 150 &  Post-fix & 07\_0189\_38 \\
\hline
\enddata
\tablenotetext{}{\footnotesize $^{*}$Sky angle refers to the orientation of the field measured in degrees east of celestial north. All angles in this table are measured with respect to the celestial frame rather than the galactic frame to be consistent with the convention of the data provided by the observatory.\\
$^{\dagger}$Chop angle refers to the direction of the chop throw, which is also the direction of PSF elongation in the fields affected by the secondary mirror issue described in text. \\
$^{\ddagger}$Data quality comments refer to the chopper mirror issue described in Section \ref{subsec:FORCASTobs7}. Data labeled `Affected' have noticeable PSF elongation in the direction of the Chop Angle. Data labeled `Nominal' or `Post-Fix' are unaffected by the mirror issue. \label{tab:tab1}}
\end{deluxetable*}

During the flight series there was an issue with the secondary chopper mirror on the telescope which caused sources to be elongated in the direction of the chop throw. Consequently, in Table \ref{tab:tab1} we have included information regarding the extent to which each field was affected by the secondary mirror issue. Fields where the elongation is known to be an issue are labeled `affected.' Fields that may also have been impacted, but only slightly are labeled `nominal', while those taken after the issue was corrected are labeled `post-fix.' In the affected data, PSF elipticities range between $\sim$ 0.1-0.4; however, several of the fields contain few, and in some cases no suitable PSF reference stars which makes measuring the data quality effects challenging. Using a deconvolution method with a suitable PSF model should be able to correct the PSF variation to an extent, although this is a complex process given level of PSF variation in the `affected' fields. We are planning to produce a set of corrected maps with a more uniform PSF to be made public with a future data release.

\subsection{SOFIA/FORCAST Observations from Prior Cycles}\label{subsec:PriorFORCASTobs}

\begin{sidewaysfigure*}[ht]
\centering
\vspace*{+2.8in}\includegraphics[width=250mm,scale=1.0]{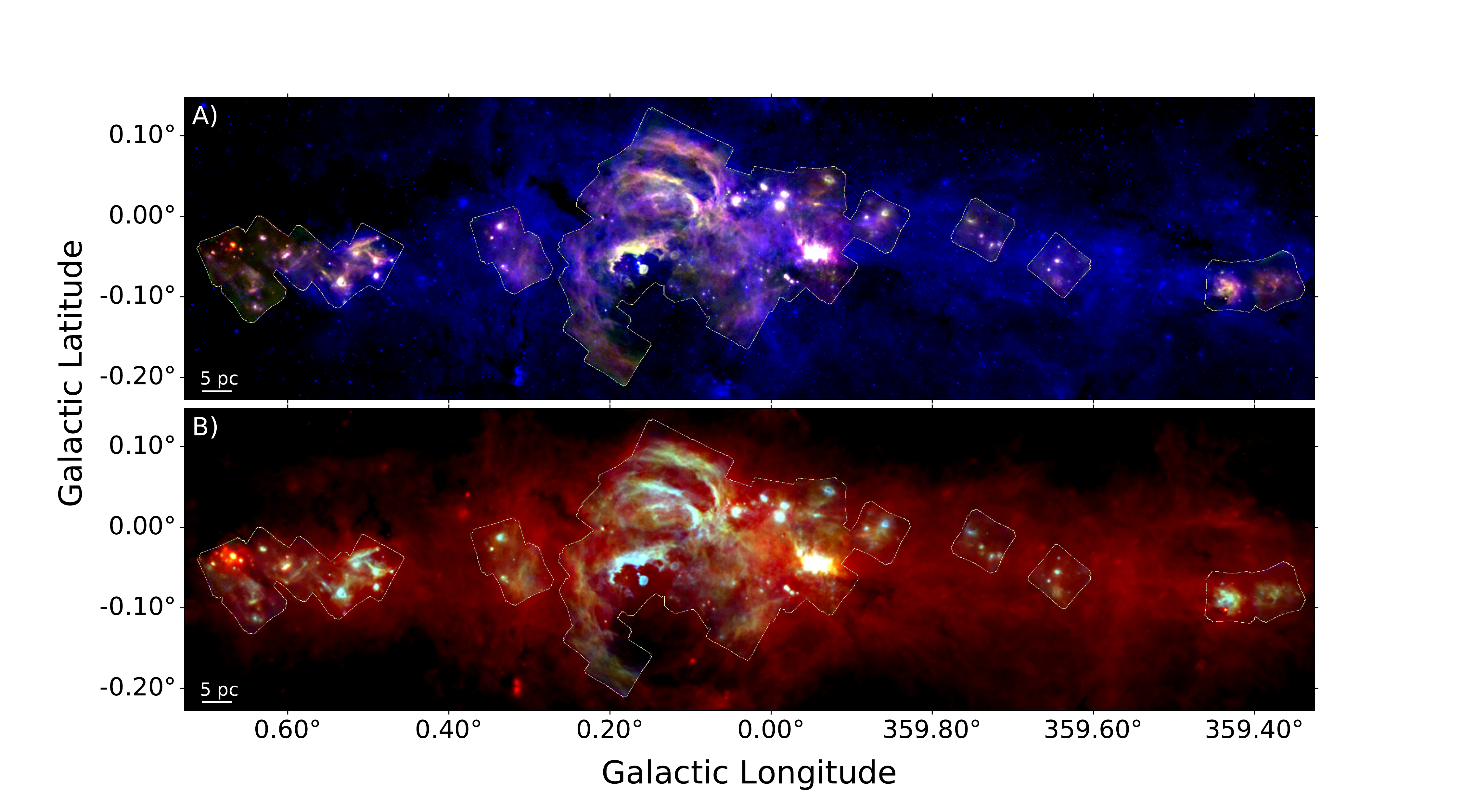}
\hspace*{-0.4in}\caption{{\footnotesize Comparison of the \textit{SOFIA}/FORCAST mosaics with other infrared survey maps. A) A false-color map of the survey region using \textit{Spitzer}/IRAC 8 $\mu$m (blue), \textit{SOFIA}/FORCAST 25 $\mu$m (green), and 37 $\mu$m (red). B) A second false-color map of the survey region using \textit{SOFIA}/FORCAST 25 $\mu$m (blue), 37 $\mu$m (green), and \textit{Herschel}/PACS 70 $\mu$m (red). Both figures show an outline of the \textit{SOFIA}/FORCAST survey footprint in white.}}
\label{fig:fig2}
\end{sidewaysfigure*}

In addition to the observations carried out in \textit{SOFIA} cycle 7, suitable data from earlier cycles were considered for inclusion in our GC survey. Several guaranteed time observations (GTO) with FORCAST were focused on targets within the GC, including the Circumnuclear Ring \citep{Lau2013},  the Sickle \ion{H}{2} region \citep{Lau2014,Hankins2016}, and the Sgr A East \ion{H}{2} regions \citep{Lau2014SgrAEastHII}. 

For the survey mosaic presented in this work, we only included observations taken in \textit{SOFIA} cycle 3 or later which consists of the Arched Filaments \ion{H}{2} region \citep{Hankins2017} and the H \ion{H}{2} regions \citep{Hankins2019}. Observation details for these data can be found in the above referenced works and a description of their incorporation in our survey mosaic can be found in the following subsection. 

\subsection{Data Processing}\label{subsec:DataProcess}

Observations were processed using the pipeline steps described in \cite{Herter2013}. The Level 3 processed data products were downloaded from the SOFIA Data Cycle System (DCS).\footnote{https://dcs.arc.nasa.gov} Images from each individual pointing were mosaicked using the \textit{SOFIA} Data Pipeline software REDUX \citep{Clarke2015} in order to construct the preliminary FORCAST Level 4 imaging mosaics that are presented in this work. Both the Level 3 and 4 data products from this program are available for download via the SOFIA DCS and the NASA/IPAC Infrared Science Archive (IRSA).\footnote{https://irsa.ipac.caltech.edu/frontpage/}

Creating the mosaic for this data set resulted in several challenges which were addressed as follows. First, observations using FORCAST on bright sources cause a negative signal offset throughout the detector that lowers the background flux levels. To correct for this, background levels were subtracted on a field by field basis to bring the overall background values to approximately zero. Next, as previously discussed, chopping and nodding requires off fields which are free of detectable mid-IR emission in order to properly subtract sky and telescope backgrounds. In cases where off fields contained sources, this results in negative point sources or regions in the data. To correct for these negative artifacts, we modeled point sources using a 2D Gaussian and if they achieved sufficient SNR ($\gtrsim$5) the source model was subtracted from the data, leaving behind the fitted background at the location of the artifact. A summary of the locations of removed sources can be found in Table \ref{tab:tab2}.

\begin{deluxetable}{cccc}
\tablecaption{Known Data Artifacts Removed in Post-Processing}
\tablehead{
\colhead{Location (l,b)} & \colhead{Filter(s)} &\colhead{Type$^*$} & \colhead{Comment$^{\dagger}$}
}
\startdata
(0.103, -0.086)   & Both & point & \ \\   
(0.098, -0.079)   & Both & point & \ \\   
(0.090, -0.070)   & Both & point & \ \\   
(359.933, 0.024)  & 25 \micron & point & \ \\      
(359.894, -0.008) & 25 \micron & point & dithered \\      
(359.896, -0.012) & 25 \micron & point & dithered \\      
(359.892, -0.015) & 25 \micron & point & dithered \\      
(0.053, -0.086)   & 25 \micron & point & \ \\    
(0.098, 0.055)    & 25 \micron & point & \ \\    
(0.103, 0.074)    & 25 \micron & point & \ \\   
(0.126, 0.111)    & Both & point & \ \\   
(0.115, 0.081)    & 25 \micron & point & \ \\   
(0.482, -0.077)   & Both & point & nearby source \\  
\hline
\enddata
\tablenotetext{}{\footnotesize $^{*}$Specifies if the artifact is point-like or extended.\\
$^{\dagger}$Comments about image artifacts. The `dithered' artifact is the same source appearing in multiple locations due to the dither pattern. The `nearby source' label denotes a nearby positive source where the photometry may be impacted by the artifact. \label{tab:tab2}}
\end{deluxetable}

Next, the issue that required the most effort to correct involved astrometry. Telescope pointing with FORCAST is only accurate to within a few pixels ($\sim$arc-seconds), therefore astrometry was absolutely calibrated using the available \textit{Spitzer} and \textit{MSX} data by matching up the centroids of point sources in common between those maps and the \textit{SOFIA} data. Precise alignment was difficult in the case of aligning to the lower resolution \textit{MSX} 21 \micron\ data or the saturated regions of the \textit{Spitzer} MIPS 24 \micron\ data. Although, a handful of bright 8 \micron\ point sources were also detected at 25 \micron\ and aided in the alignment of several of fields.

Slight changes in the focal plane distortion across the array and limited calibration data, also contributed to some elongation and smearing of sources on scales similar to the elongation caused by the secondary chopping mirror issue as discussed above. This distortion is variable across the array with the worst effects at the edges resulting in some misalignment of up to a few pixels ($\sim$arc-seconds). Based on these issues we estimate the astrometry of the final \textit{SOFIA} mosaic is at worst 3 pixels or $\sim$ 2$\arcsec$. However, comparing common sources between the FORCAST and \textit{Spitzer} suggests the FORCAST mosaic astrometry is typically better than 1 pixel.\footnote{The median offset for a modest-sized sample (N=25) of randomly selected sources between the data sets is $\sim$0.4 pixels or $\sim$0.3 $\arcsec$.} More calibration data are expected to be taken in the next SOFIA cycle, in which case the overall astrometric precision of the maps may be improved in a future data release.

Finally, changes to the data reduction pipeline between \textit{SOFIA} cycle 7 and earlier cycles presented several issues for combining the various data sets into the mosaics presented in this work. These issues were largely overcome by reprocessing the earlier cycle 3 and 4 data using REDUX with modifications to improve backwards compatibility. While we present initial mosaics produced as part of this program, there is still work needed to optimally combine the older and newer data sets. For example, the new processing for the 37 \micron\ observations has resulted in a 3-pixel wide gap in a small section between two fields in the Arched Filaments near (0.116, 0.073), and there is a similar 1-pixel wide gap in a section between another set of fields in the 25 \micron\ map located near (0.148, 0.065). Neither of these gaps where present in earlier versions of the data at the time they were processed in cycles 3 and 4 and may point to an issue with differences in distortion correction over time. For the present data release, we have used neighboring pixels to interpolate over these `missing' pixels in the small gap areas. As part of a future data release, we are planning updates to the REDUX package that will help with these and other issues in order to improve future versions of the survey mosaics.

\subsection{Additional Data From The Literature} \label{subsec:otherdata}

In our initial study of the \textit{SOFIA}/FORCAST mosaics we compared the maps with a number of other prominent GC surveys from the literature including the \textit{HST} Paschen-$\mathrm{\alpha}$ Survey of the Galactic Center \citep{Wang2010}, the GLIMPSE \textit{Spitzer}/IRAC survey \citep{Churchwell2009}, and the \textit{Herschel} Hi-GAL survey \citep{Molinari2010}. These data sets provide a number of useful comparisons that we will discuss in subsequent sections. We also referenced maps and data from a number of other works in the literature which were focused on individual regions within the mosaics, and these are discussed in the relevant sections in the text.

\section{Discussion} \label{sec:Discussion}

\subsection{Comparison with other IR surveys}

The GC is probably one of the most surveyed portions of sky and a number of high quality data sets exist at nearly all available wavelengths. In this section, we focus on comparisons between the FORCAST mosaics and earlier survey maps produced by \textit{Spitzer}/IRAC at 8 $\mu$m and \textit{Herschel}/PACS at 70 $\mu$m. Three-color combinations of these data sets can be found in Figure \ref{fig:fig2}. 

The morphology of the 8 $\mu$m emission contains numerous similarities, but also important differences when compared to the 25 and 37 $\mu$m emission. The origin of the 8 $\mu$m emission differs from that at 25 and 37 $\mu$m; 8 $\mu$m emission primarily traces very small transiently heated grains, particularly Polycyclic Aromatic Hydrocarbons (PAHs) which have prominent emission features in this wavelength region, while longer wavelength emission traces emission from larger, presumably silicate grains. Throughout the region, there is a diffuse 8 $\mu$m component corresponding to most of the IR bright regions at 25 and 37 $\mu$m, with the 8 $\mu$m emission appearing more extended in comparison. There are notable exceptions to this trend, however, which likely points to variations in extinction over the survey area which can impact our observations even at longer wavelengths. For example, Sgr B2 is relatively dark at 8 $\mu$m compared to many other regions within the survey, and sources in this region of the FORCAST data have very red colors. Both of these qualitative indicators suggest high extinction, and in the case of Sgr B2 it is already well established this region suffers from significant local extinction \citep[e.g.,][]{Scoville1975}.

\begin{figure*}[ht]
\centering
\includegraphics[width=150mm,scale=0.5]{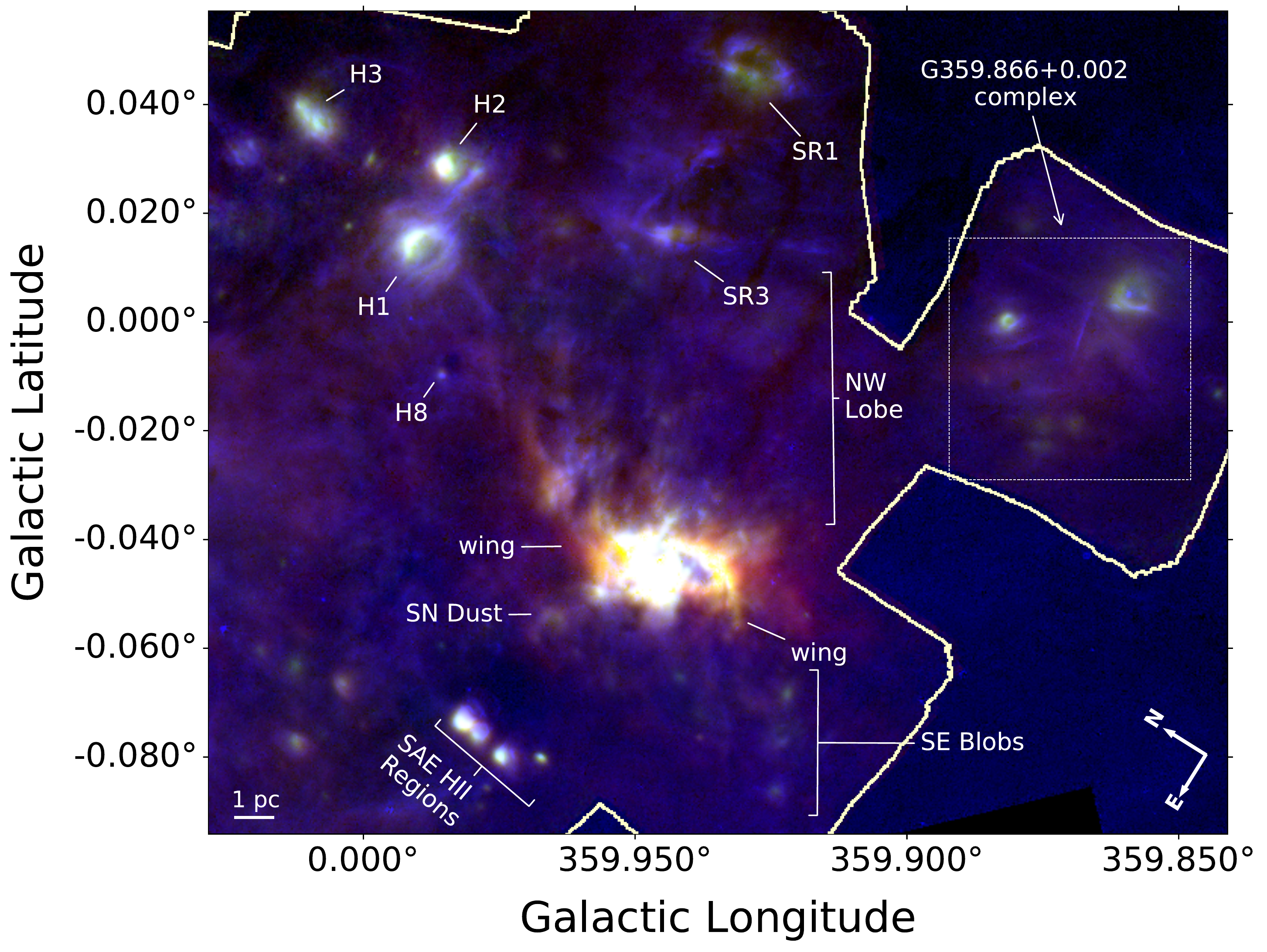}
\caption{{\footnotesize A false color map of the Sgr A complex created with the \textit{HST} Paschen-$\alpha$ emission (blue), \textit{SOFIA}/FORCAST 25 $\mu$m (green), and 37 $\mu$m (red) data. Several of the features discussed in text are labeled for reference. Sgr A East is abbreviated `SAE' for brevity in the figure label. The dashed box shows the region around the G359.866+0.002 complex which is featured in Figure \ref{fig:G359+002} and an outline of the FORCAST survey footprint is also overlaid as a solid white outline.}}
\label{fig:fig3}
\end{figure*}

Comparison of the 70 $\mu$m emission and the 25 and 37 $\mu$m emission highlights a number of notable features. The emission at each of these wavelengths is primarily thermal continuum with the 25 $\mu$m emission tracing relatively warm $\sim$ 100 K dust and the 70 $\mu$m emission tracing cooler $\sim$40 K dust. Many of the \ion{H}{2} regions throughout the GC contain significant emission from warm dust which results in bright emission at 25 $\mu$m (Figure \ref{fig:fig2}). This highlights the importance of the 25 $\mu$m data when studying recent star formation both in giant \ion{H}{2} regions like the Arched Filaments and Sickle as well as well as in smaller \ion{H}{2} regions which may be associated with individual massive stars. Furthermore, the spatial resolution of FORCAST enables us to resolve many `bubble-like' structures down to a physical scale of $\sim$0.1 pc. For isolated sources, this is sufficient to resolve all but the most compact \ion{H}{2} regions at the GC distance.

Another striking feature of the 25, 37, and 70 $\mu$m data in Figure \ref{fig:fig2} is the number of compact red sources spread throughout the region. Prime examples of this can be seen in both Sgr B2 and Sgr C. In particular Sgr B2 contains several of the most luminous 70 $\mu$m sources in the GC (see \S\ref{sec:SgrB}). In the FORCAST data we clearly observe these same objects with strong emission at 37 $\mu$m, suggesting deeply embedded star formation.

\subsection{The Sgr A Complex}\label{sec:SgrA}

One of the most interesting regions within the FORCAST mosaics is the Sgr A complex. Some of the earliest FORCAST observations of the GC were focused on the Circumnuclear Disk \citep[CND; ][]{Lau2013}; but the limited field of view prevented study of any large extended structures that might be associated. Our survey significantly expands the coverage of this region allowing us to examine various warm dust features extending from the position of the CND out to several parsecs. Among the highlights from this region is the mapping out of mid-infrared emission at the position of the NW X-ray lobe, extending from the CND \citep{Ponti2015}. In addition, we observe several other extended structures which have been noted in radio and ionized gas observations the region, enabling us to compare the warm dust and ionized gas throughout the region \citep[e.g.,][]{Zhao2016}. 

\begin{figure*}[ht]
\centering
\includegraphics[width=145mm,scale=0.5]{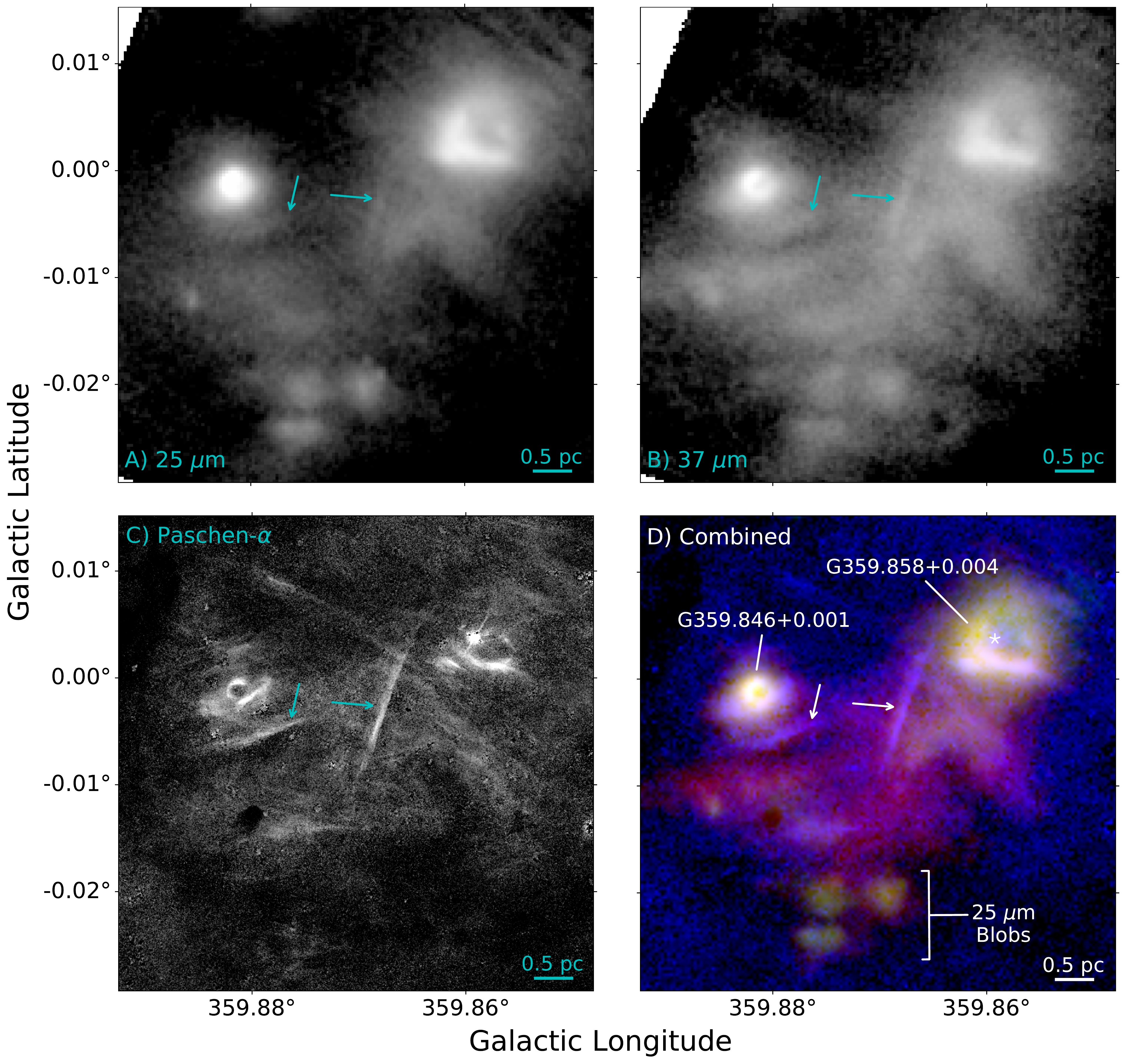}
\caption{{\footnotesize A) \textit{SOFIA}/FORCAST 25 $\mu$m data of the G359.866+0.002 complex. Two notable linear features are marked with arrows both here and in the following panels. B) \textit{SOFIA}/FORCAST 37 $\mu$m data of the G359.866+0.002 complex. C) \textit{HST} Paschen-$\alpha$ data of the G359.866+0.002 complex. D) A false color map of the G359.866+0.002 complex created with the Paschen-$\alpha$ (blue), 25 $\mu$m continuum (green), and 37 $\mu$m continuum (red) data. Additional labels are provided for the two dusty nebulae discussed in text and the location of the emission line star 2MASS J17451618-2903156 is marked with a white star. Three extended sources to the south are labeled as the 25 $\mu$m blobs.}} 
\label{fig:G359+002}
\end{figure*}

 We note remarkably detailed structures in the warm dust emission in this region (Figure \ref{fig:fig3}). Outside of the brightest mid-IR features tracing the `minispiral' feature and the inner edge of the CND, also referred to as the Circumnuclear Ring (CNR), we observe fainter extended structures protruding from the CNR which correspond to the well-known ``wing'' features \citep[often referred to as the NW and SE Wings; ][]{Zhao2016}. Similar streamers are also detected in this region in molecular gas \citep{Liu2012}, and the emission observed by FORCAST may be related to these molecular clouds. Dust near the NW wing appears to extend for several arcminutes, terminating near the position of the source H1 in the H \ion{H}{2} regions. This structure also appears to trace the edge of the NW X-ray lobe, although more detailed analysis is needed to understand possible interaction between the x-ray emitting region and the dusty infrared clouds. 
 
 At the position of the NW lobe we find relatively faint, extended mid-IR emission that extends northward to a few relatively bright 25 $\mu$m objects. These 25 $\mu$m sources appear to be associated with the `smoke rings' discussed in \cite{Zhao2016}. In the infrared, only one of these sources has a ring-like appearance (smoke ring \#1; SR1 in Figure \ref{fig:fig3}), and might be associated with the nearby massive star CXOGC J174516.7-285824 \citep{Mauerhan2010}. The IR emission of the third smoke ring (SR3) appears to `fill in' the ring-like structure observed in radio and ionized gas, and suggests this feature may simply be a portion of a molecular cloud with prominent ionized edges. The second smoke ring structure from \cite{Zhao2016} has little to no detectable emission in the mid-infrared wavelengths presented in this work.  
 
To the east of the CNR we see the dust emission associated with the Sgr A East supernova remnant which was reported in \cite{Lau2015}. To the south of the CND there are a number of relatively faint sources and structures. The location of several of these features appear to correspond to the location of the `SE blobs' reported by \cite{Zhao2016} and may be related to disrupted cloudlets in the region \cite[see Figure 7 from][]{Liu2012}. The origin of material in this region is of particular interest because it could signify outflow from the CND \citep[e.g., ][]{Wang2010}. We will explore these and other IR features in greater detail in a future paper focusing on Sgr A. 


The \ion{H}{2} region complex G359.866+0.002 (Figure \ref{fig:fig3}) lies to the equatorial southwest of the CND with a projected separation of $\sim$2.5' ($\sim$6 pc). While this complex is not typically discussed in the context of Sgr A, this region contains a number of interesting features and is the southernmost portion of our coverage of Sgr A. G359.866+0.002 was found as a collection of Paschen-$\alpha$ emitting sources in \cite{Wang2010} where it was featured for its linear ionized gas features. In the FORCAST data, we observe dust emission at 25 and 37 $\mu$m that traces many of these same narrow ionized gas features (Figure \ref{fig:G359+002}). These structures could point to locally strong magnetic fields or shocks that are sculpting the gas and dust emission. We also note two extended nebulae in the complex that are prominent at 25 $\mu$m, indicating dust that is likely being heated by luminous nearby stars. The 25 $\mu$m nebulae G359.858+0.004 may be associated with the nearby emission line star 2MASS J17451618-2903156 from \citep{Mauerhan2010}, while the other 25 $\mu$m source (G359.846+0.001) is cataloged as a YSO candidate based on ISOGAL data \citep{Immer2012ISOgal}. Although the latter source is not present in the YSO catalog published in \citet{Yusef-Zadeh2009} and was not observed as part of the \textit{Spitzer}/IRS sample of Galactic center YSOs presented in \citet{An2011}. To the south of these sources there are a collection of dusty ridges visible at 37 \micron\, some of which have associated ionized gas emission and others which do not. Further to the south there is a collection of interesting extended 25 \micron\ sources, which we refer to as 25 \micron\ `blobs'. As part of a future data release for this program, we will produce source catalogs for the 25 and 37 $\mu$m data sets and publish additional information on possible YSO candidates determined from infrared colors of the \textit{SOFIA} data.

\subsection{The Arched Filaments and Sickle \ion{H}{2} Regions} \label{sec:ArchesSickle}

\begin{figure*}[ht]
\centering
\includegraphics[width=150mm,scale=1.0]{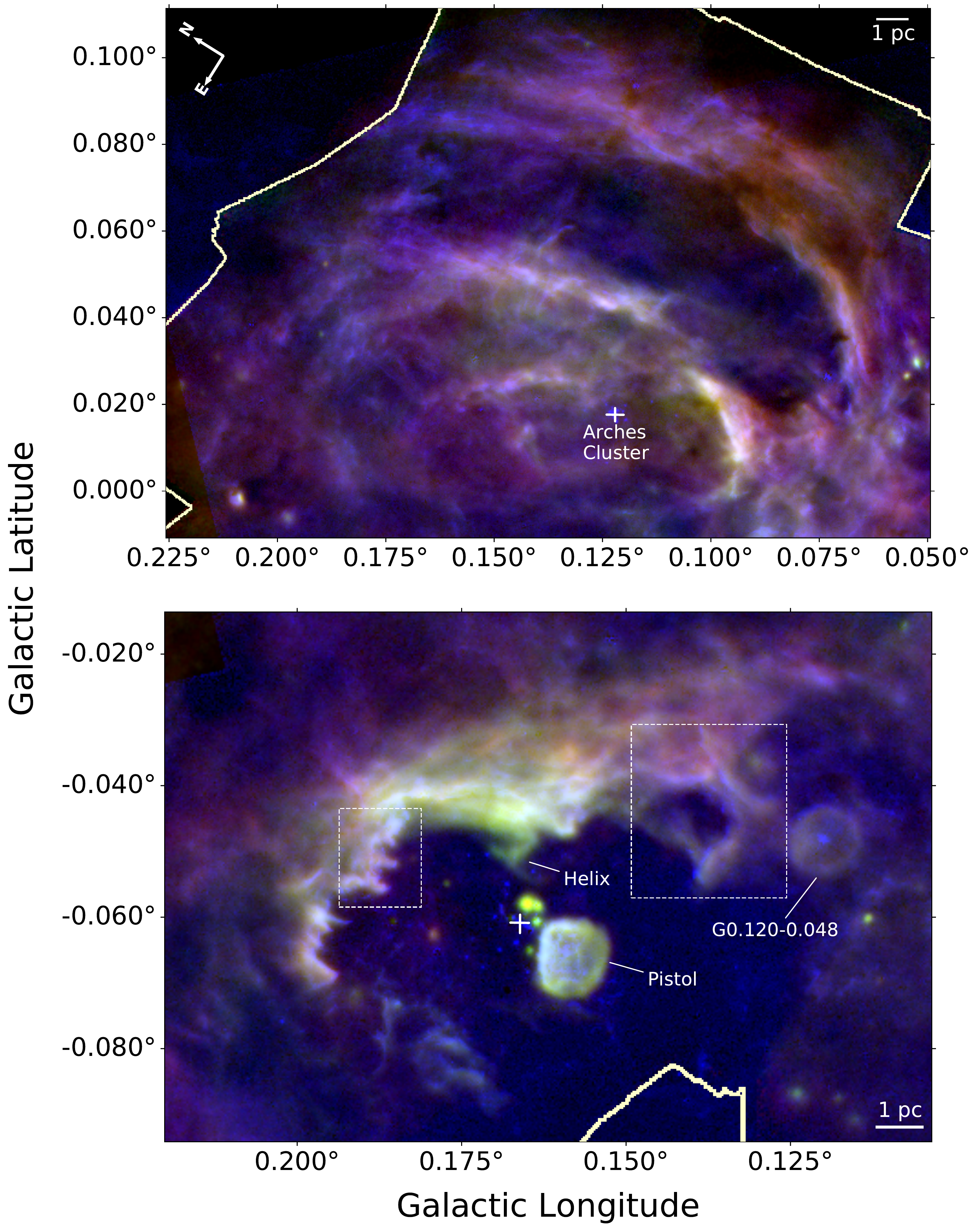}
\caption{{\footnotesize A false-color map of the Arched Filaments (Top) and the Sickle \ion{H}{2} Region (Bottom) created with the \textit{HST} Paschen-$\alpha$ emission (blue), \textit{SOFIA}/FORCAST 25 $\mu$m (green), and 37 $\mu$m (red) data. The approximate locations of the Arches and Quintuplet clusters are marked with a white crosshair. Several infrared sources associated with the Quintuplet cluster can be seen near the marker, while the Arches cluster has a distinct lack of bright mid-infrared counterparts. Additional sources discussed in the text are labeled for reference, while the two dashed boxes indicate regions that are featured in a subsequent Figure. An outline of the FORCAST survey footprint is also overlaid as a solid white outline in both plots. Previous FORCAST observations of these regions have been featured in earlier works \citep{Lau2016Helix,Hankins2016,Hankins2017} which provide more detailed analysis.}}
\label{fig:fig5}
\end{figure*}

\begin{figure*}[ht]
\centering
\includegraphics[width=150mm,scale=1.0]{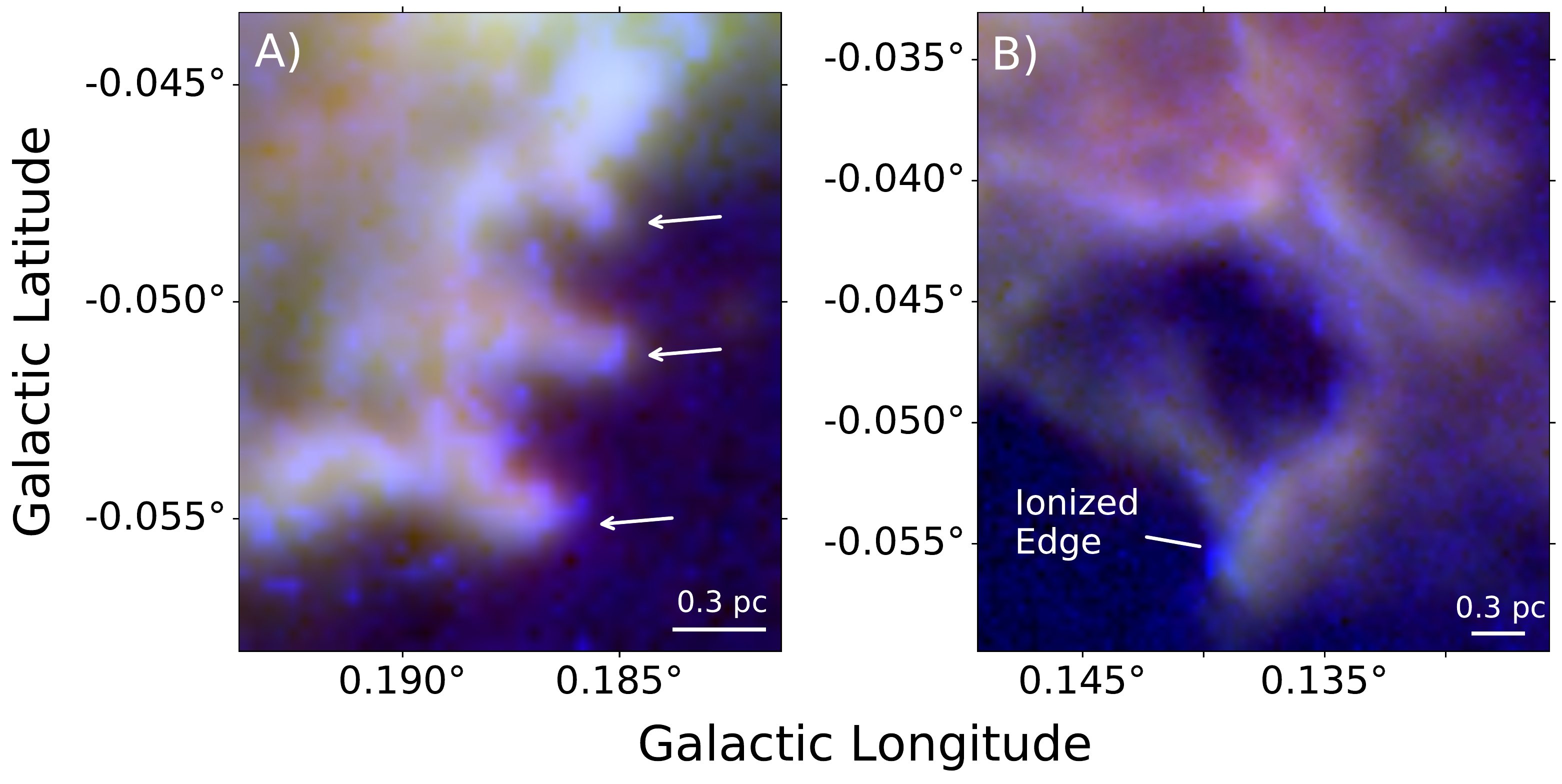}
\caption{{\footnotesize A `zoom-in' on two interesting regions within the Sickle \ion{H}{2} region. Panel A) shows several of the finger-like structures in the `blade' region. Arrows point out a few of the more prominent of these features which appear to measure $\sim$0.1--0.3 pc in size. Panel B) shows the `ladder' region which shows prominent ionized gas emission (blue) on the edges of the dusty clouds. Both panels were created using the same 3-color map as Figure \ref{fig:fig5}.}}
\label{fig:sicklezoom}
\end{figure*}

The Arched Filaments and Sickle \ion{H}{2} regions are some of the most prominent structures in the GC at both infrared and radio wavelengths \citep[e.g.,][]{Yusef-Zadeh1987}. Earlier work has been presented on \textit{SOFIA}/FORCAST observations of both of these regions \cite[e.g.,][]{Lau2014,Lau2016Helix,Hankins2016,Hankins2017}, however, our discussion of the present GC survey would be incomplete without some mention of these important regions. In this section we highlight how the combined archival and new survey data can improve our picture of the physical processes at work in the Arched filaments and Sickle (Figure \ref{fig:fig5})

\begin{figure*}[ht]
\centering
\includegraphics[width=150mm,scale=1.0]{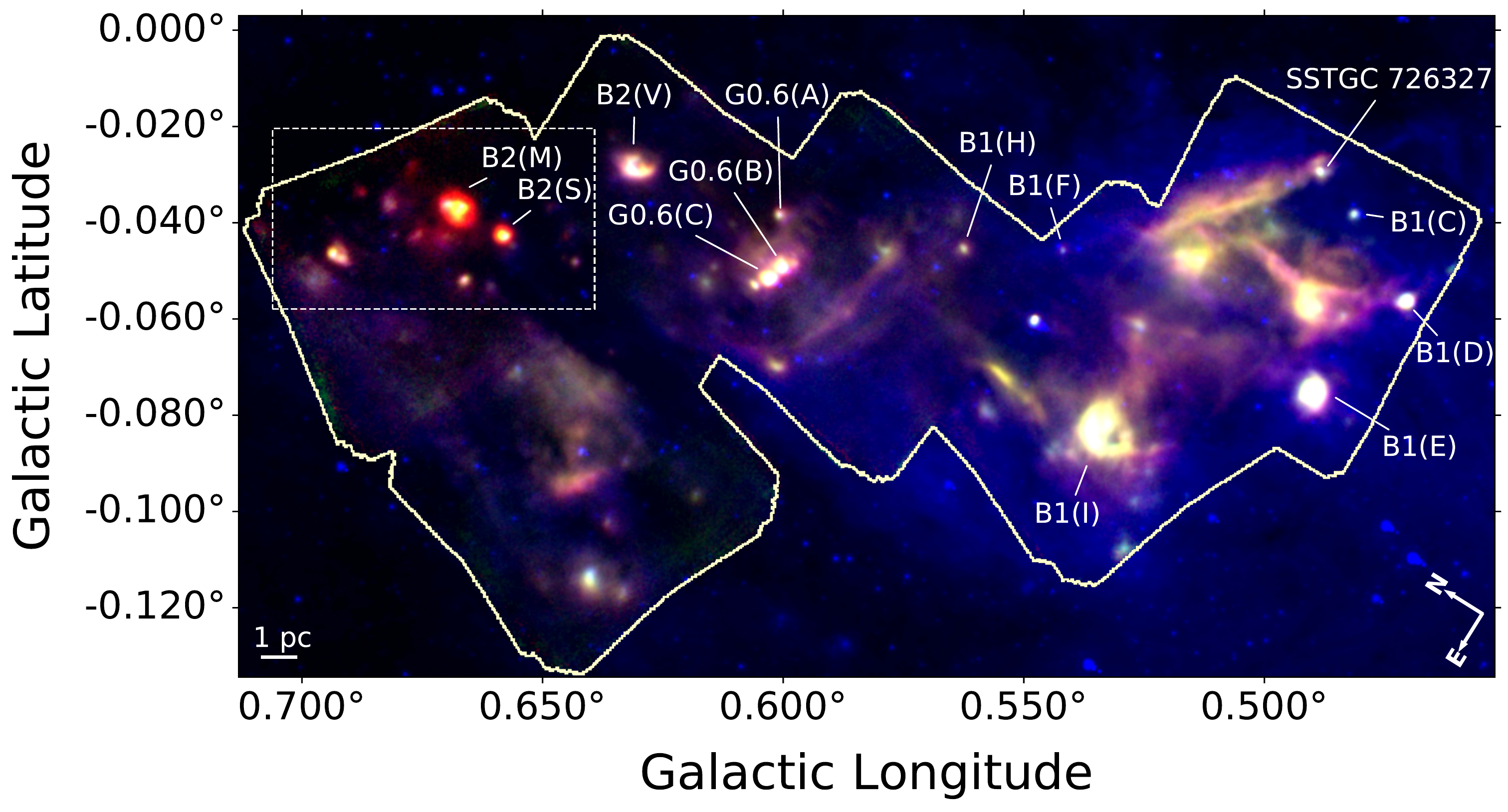}
\caption{{\footnotesize A false-color map of the Sgr B complex using \textit{Spitzer}/IRAC 8 $\mu$m (blue), \textit{SOFIA}/FORCAST 25 $\mu$m (green), and 37 $\mu$m (red) data. We use the IRAC 8 $\mu$m data in this figure rather than the \textit{HST} Paschen-$\alpha$ survey because the Sgr B region is outside of the \textit{HST} survey footprint. Sources of interest are labeled throughout the region following the naming convention of \citet{Mehringer1992,Mehringer1993} with the exception of Sgr B2 Main, Sgr B2 South, and Sgr B1(A) which is associated with the massive YSO SSTGC 726327. Labels for sources in the G0.6-0.0 region are abbreviated `G0.6'. The dashed box shows a region of the map that is featured in Figure \ref{fig:B2zoom}, and the footprint of the \textit{SOFIA}/FORCAST survey is also shown as a solid white outline for reference.}}
\label{fig:fig6}
\end{figure*}

The Arched Filaments complex is a collection of filamentary features that have a distinctive arched morphology due to our viewing geometry of the system \citep{Lang2001,Lang2002}. These structures are well known for their relation to the non-thermal emission in the GC Radio Arc, which meets with and undergoes a brightness discontinuity at the northern edge of the Arched Filaments \citep{Lang2001} and appears to travel through the position of the Sickle \ion{H}{2} region \citep{Lang1997}. Dust temperatures within the Arched Filaments \ion{H}{2} region are consistent with the Arches cluster being the primary heating source \citep{Hankins2017}, as well as the primary source of ionizing radiation \citep[e.g.][]{Lang2001}. Our previous FORCAST study of this region did not reveal any signs of ongoing star formation within the infrared bright filament structures \citep[Figure \ref{fig:fig5} upper panel;][]{Hankins2017}. Furthermore, we do not detect any infrared point sources near the location of the Arches cluster at 25 or 37 $\mu$m which points to a lack of evolved, dust forming stars due to its relatively young age \citep[$\sim$2-3 Myr;][]{Figer1999,Stolte2002}.

\begin{figure*}[ht]
\centering
\includegraphics[width=130mm,scale=1.0]{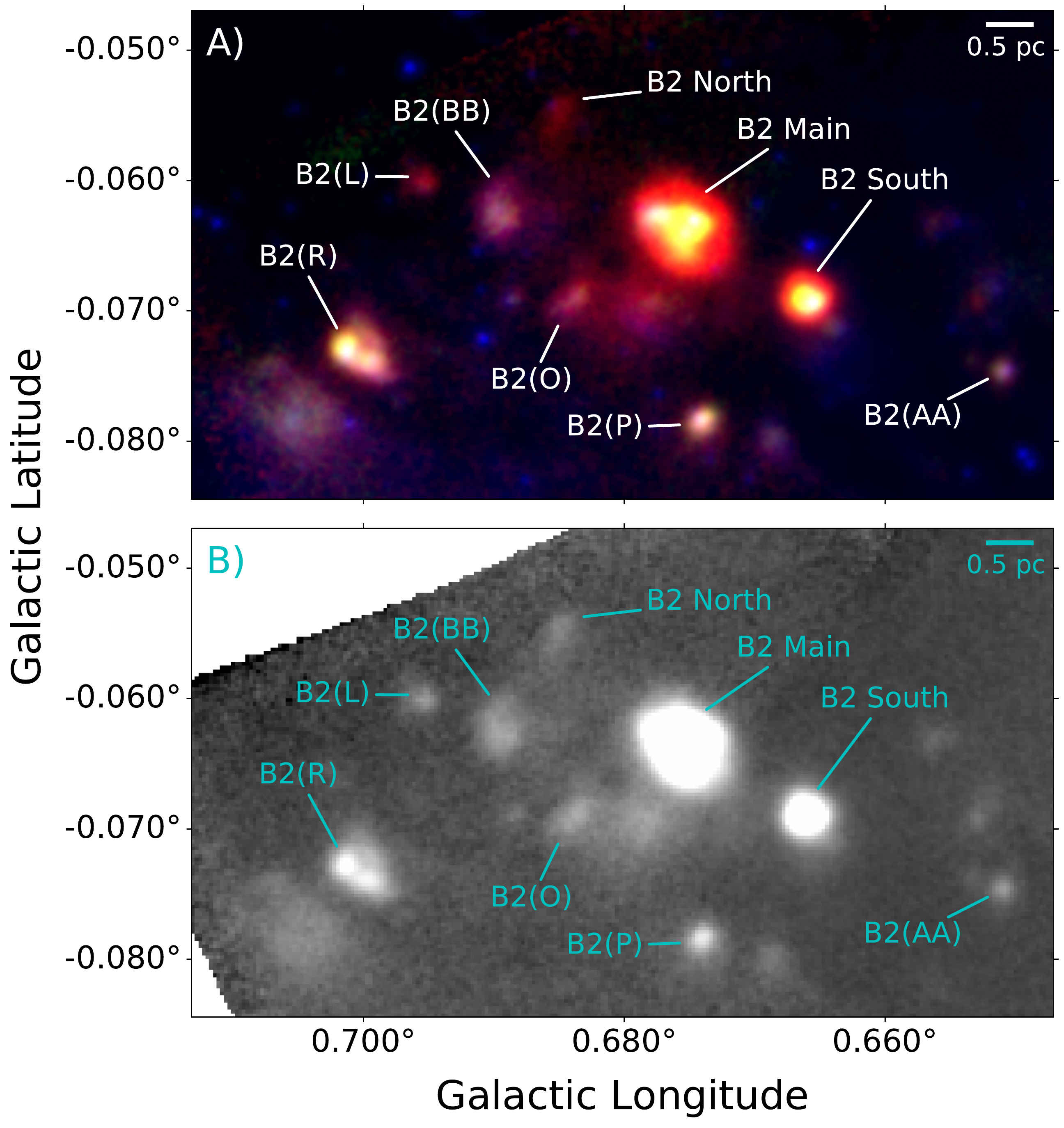}
\caption{{\footnotesize A `zoom-in' on Sgr B2 with various sources of interest labeled following the naming convention from \citet{Mehringer1993} with the exception of Sgr B2 Main, North, and South. Panel A) was created using the same three-color map from Figure \ref{fig:fig6}. Panel B) shows the same area with only the 37 $\mu$m data plotted in grayscale.}}
\label{fig:B2zoom}
\end{figure*}

The Sickle \ion{H}{2} region has a number of interesting features as seen both in the morphology of the Sickle proper, and in the numerous infrared sources primarily related to the Quintuplet cluster. First, the northern portion of the Sickle cloud (the `blade') shows several `finger'-like structures (Figure \ref{fig:fig5} lower panel and Figure \ref{fig:sicklezoom}) reminiscent of features in M16 \citep{Cotera2006}. It is unclear if these structures are actively star forming like their M16 counterparts, although we may simply lack sufficient sensitivity and spatial resolution to adequately determine this at the GC distance. The morphological differences between the northern and western parts of the Sickle suggests they are impacted by the magnetic field organization within the cloud. Notably the perpendicular part of the cloud (the `handle') does not display similar `fingers'. Although, there is a relatively faint, wispy, helical structure protruding from the `handle' which \citet{Lau2016Helix} suggest is related to outflow from a massive star that appears near the cloud edge. 

Gas and dust within the Sickle are most likely ionized and heated by the highly luminous Quintuplet cluster \citep[$L=3\times10^7~ L_{\odot}$;][]{Figer1999}. Unlike the Arches cluster, the Quintuplet cluster contains five luminous infrared sources from which the cluster name is derived \citep{Okuda1990,Nagata1990}. These five `Quintuplet Proper Members' are likely Wolf-Rayet stars which are undergoing active dust production \citep{Tuthill2006,Najarro2017}, and represent only a few of the massive stars belonging to the cluster (Figure \ref{fig:fig5} lower panel). Our survey observations also show dust associated with two extended nebulae surrounding the candidate Luminous Blue Variable (LBV) stars, the Pistol star and G0.120-0.048 \citep{Lau2014}. While there is a third known LBV candidate in this region, qF362 \citep{Figer1999}, there is no obvious infrared counterpart associated with this source in the FORCAST maps. The data obtained as part of our survey of the Sickle region will be explored in depth in a subsequent paper focused on the `finger'-like structures and other prominent features of the \ion{H}{2} region discussed in this section.


\subsection{The Sgr B1 \& Sgr B2 Complexes} \label{sec:SgrB}

The Sgr B complex is the easternmost region mapped in our survey. Sgr B is located along the galactic plane with a projected separation between $\sim$16'--23' ($\sim$40--50pc) from Sgr A as measured from the near and far side of Sgr B. This complex is most frequently discussed in terms of three distinct parts: Sgr B1, Sgr B2, and G0.6-0.0 which lies between Sgr B1 and Sgr B2 (Figure \ref{fig:fig6}). Sgr B2 is one of the most massive and active star forming regions in our Galaxy with a few hundred sources which are likely a mix of young stellar objects and slightly more developed stars that have produced \ion{H}{2} regions \citep{dePree1995,dePree1996,Ginsburg2018}. In contrast, Sgr B1 is a more evolved star-forming region kinematically linked to Sgr B2, which contains a population of less extinct (embedded) \ion{H}{2} regions \citep{Mehringer1992,Mehringer1993,Lang2010}. There is, however, recent evidence to suggest that these \ion{H}{2} regions are a result of evolved massive stars passing through the dense medium in Sgr B1, as opposed to young O/B-stars that have formed within the cloud \citep{Simpson2018SgrB1}.

\begin{figure*}[ht]
\centering
\includegraphics[width=150mm,scale=1.0]{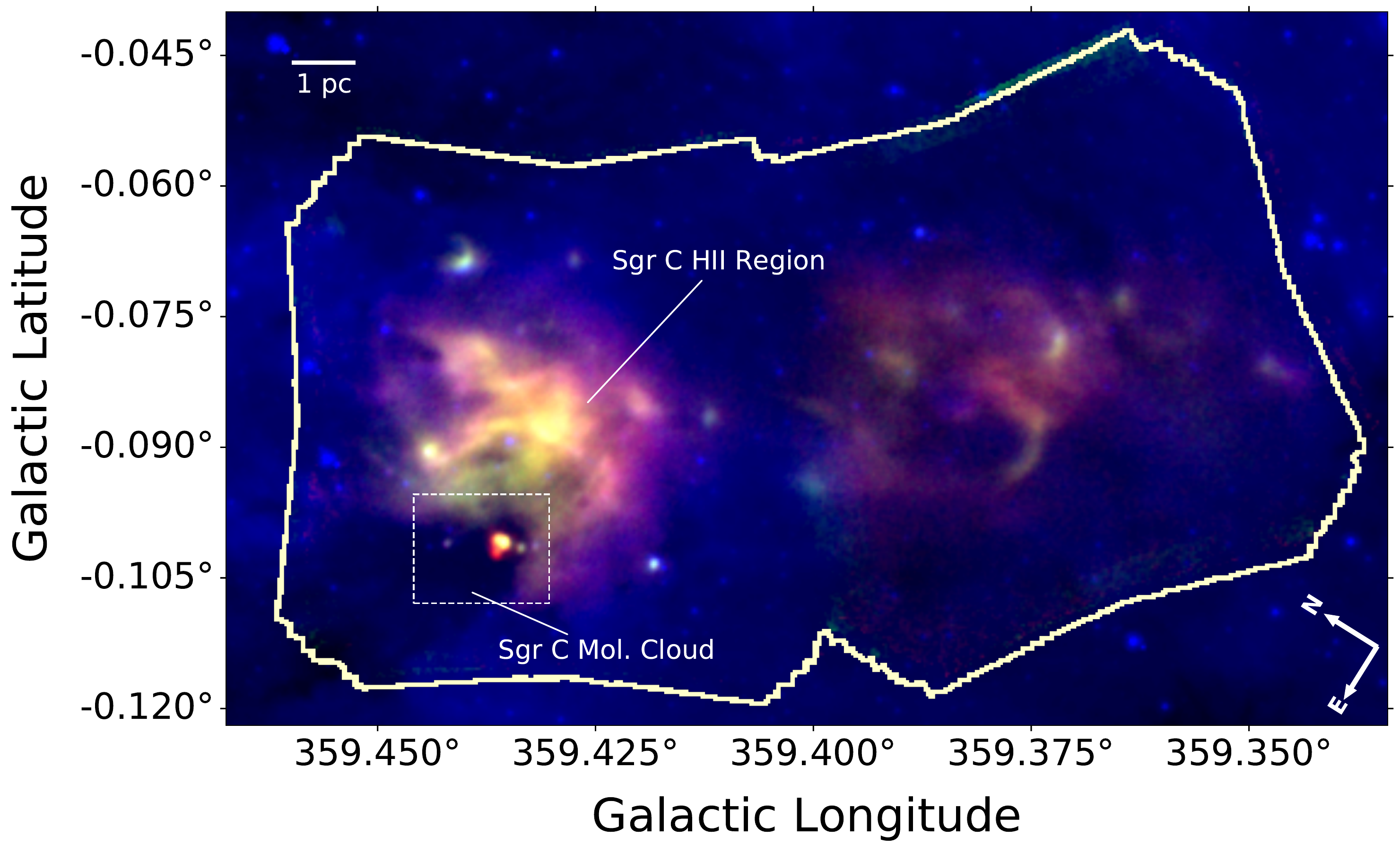}
\caption{{\footnotesize A false-color map of Sgr C using \textit{Spitzer}/IRAC 8 $\mu$m (blue), \textit{SOFIA}/FORCAST 25 $\mu$m (green), and 37 $\mu$m (red) data. We use the IRAC 8 $\mu$m data in this figure rather than the \textit{HST} Paschen-$\alpha$ survey because the Sgr C region is outside of the \textit{HST} survey footprint. A few regions of interest discussed in the text are labeled and the \textit{SOFIA}/FORCAST survey footprint is also shown for reference.}}
\label{fig:fig7}
\end{figure*}

The FORCAST observations of Sgr B reveal a number of interesting features throughout the region. Qualitatively, Sgr B1 shows considerable extended emission with large filamentary structures, shells, and bubbles most of which have radio counterparts, while Sgr B2 appears as a collection of compact reddened sources. A few of the brightest mid-infrared sources in this region include the well-known complexes Sgr B2 Main and Sgr B2 South as well as Sgr B2 V \citep[e.g.,][]{Etxaluze2013}. Near Sgr B2 Main and South we note a several relatively faint IR sources which appear to be counterparts to known compact and ultra-compact \ion{H}{2} regions (Figure \ref{fig:B2zoom}). The presence of these sources provide further evidence for recent star formation activity outside of the brightest complexes in Sgr B2, which is consistent with recent ALMA observations that show evidence of extended star formation outside of the main clusters in Sgr B2 \citep{Ginsburg2018}. Study of the brightest and faintest mid-IR sources in this region will require use of both the FORCAST data set and the \text{Spizer}/MIPS 24 $\mu$m data. As part of a future data release we are planning to create a combined map between these two datasets which will allow for study of regions like Sgr B2 where large dynamic range is needed to effectively dissect the region.

Although the FORCAST data of Sgr B1 do show a bright compact source coincident with one of the massive YSOs (SSTGC 726327) identified in \citet{An2011}, the other candidates from that paper within our observations are unremarkable. There is, however, a strong ridge of warm dust emission immediately adjacent to one of the Wolf-Rayet stars found in \citet{Mauerhan2010} that does not have a strong radio counterpart and might be evidence of the impact of a orbiting massive star not formed in situ.  \citet{Simpson2018SgrB1} suggested, based on maps of the [\ion{O}{3}] 52 and 88 \micron\ lines in Sgr B1 with FIFI-LS, that the region may not, in fact, be forming stars as we see in Sgr B2, but rather the observed emission may be the result of passing massive stars, such as is seen in the regions surrounding the Arches and Quintuplet clusters.

\subsection{Sgr C \& Neighbors} \label{sec:SgrC}

The Sgr C complex is the westernmost region mapped in our survey. Sgr C is located along the Galactic plane with a projected separation of $\sim$15' ($\sim$35 pc) from Sgr A. Our observations of Sgr C focused on the main Sgr C \ion{H}{2} region with an additional adjacent pointing to the west containing the position of source `C' from \cite{Liszt1995}. Earlier measurements of ionized and molecular gas suggest that the \ion{H}{2} region has a shell-like morphology which is likely created by massive stars that have blown out a cavity in the surrounding gas \citep{Lang2010}. The brightest portions of the extended 25 and 37 $\mu$m emission we observe in this region are associated with the Sgr C \ion{H}{2} region (Figure \ref{fig:fig7}). However, the morphology of the dust appears quite complex compared to a simple shell configuration.

\begin{figure*}[ht]
\centering
\includegraphics[width=150mm,scale=1.0]{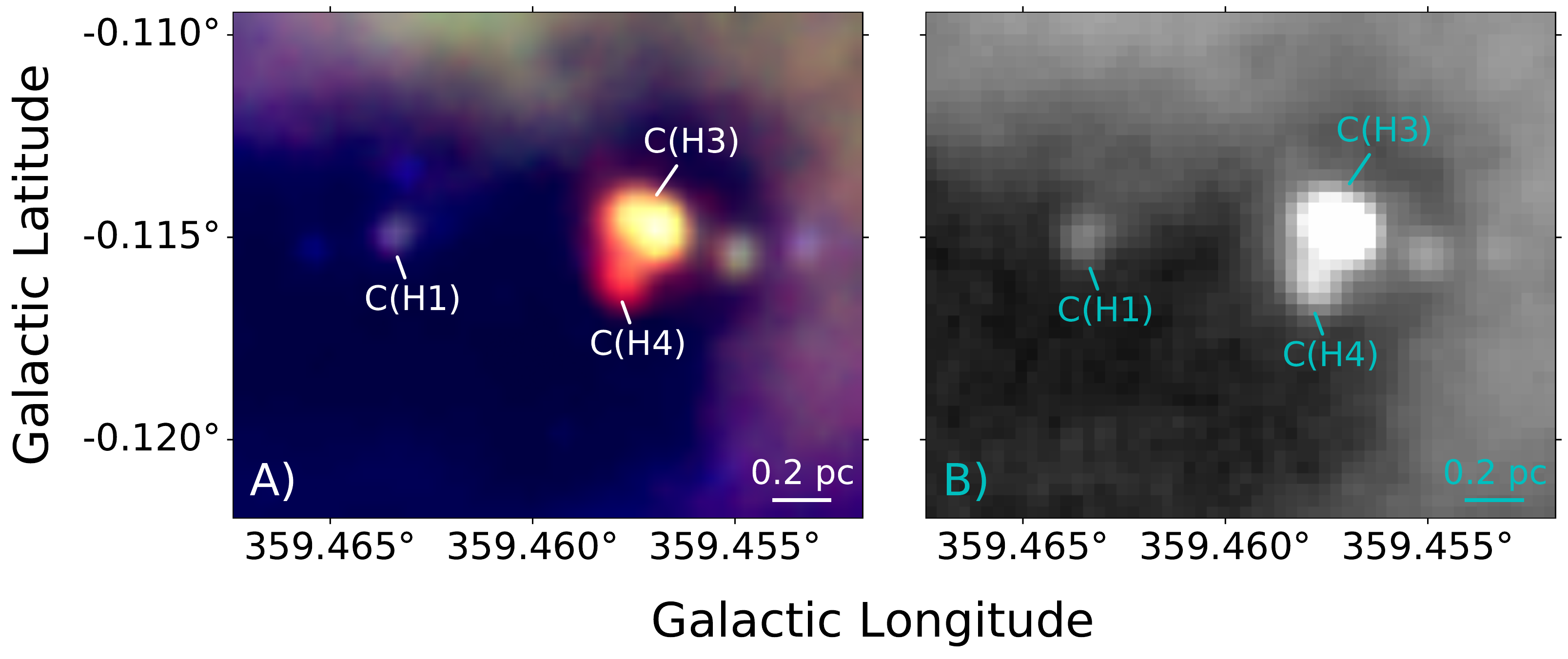}
\caption{{\footnotesize A zoom in on the Sgr C molecular cloud with sources labeled as in \cite{Lu2019}. Panel A) was created using the same three-color map as in Figure \ref{fig:fig7}. Panel B) shows the same area using the 37 $\mu$m data in grayscale.}}
\label{fig:SgrCzoom}
\end{figure*}

We also observe the Sgr C Molecular Cloud as an infrared dark cloud at 25 and 37 $\mu$m. The molecular cloud has a velocity of -55 km s$^{-1}$ \citep{Kendrew2013}, indicating that it is likely associated with the \ion{H}{2} region, which has a measured velocity of -65 km s$^{-1}$ from recombination line emission \citep{Lang2010}. Toward the dark cloud we find a number of bright mid-infrared sources (Figure \ref{fig:SgrCzoom}). The most luminous of these sources (Sgr C H3) is reminiscent of the bright, red sources we observe in Sgr B2. This object has been previously cataloged as an ultra-compact \ion{H}{2} region by \cite{Forster2000}, and \cite{Kendrew2013} show that this source has additional substructure, including two dusty protostellar cores and signatures of an outflow which are indicative of ongoing high-mass star formation. There are a two additional infrared sources near Sgr C H3, which are also cataloged as \ion{H}{2} regions in \cite{Lu2019} and can be found in Figure \ref{fig:SgrCzoom}.

The Sgr C complex is also known for its prominent non-thermal filament (NTF) \citep[e.g.,][]{Roy2003}. Our survey did not cover the location of the filament because of our focus on the mid-infrared bright regions. However, the NTF should be discussed along with the \ion{H}{2} region, because of the likely association of these features. In fact, large \ion{H}{2} regions may be key to producing bright non-thermal features at radio wavelengths by providing a vast supply of free electrons that are accelerated to relativistic velocities via locally strong magnetic fields \citep{Serabyn1994,Uchida1996}. There are numerous similarities between the well-known GC Radio Arc and the Sgr C NTF that warrant further study. This is another area where combining the MIPS and FORCAST data is needed to create a more complete picture of this region in the infrared and improve our understanding of the environments that give rise to prominent radio features. Certainly this will be interesting area to study with the release of high level data products from this survey.

\begin{figure*}[ht]
\centering
\includegraphics[width=150mm,scale=0.5]{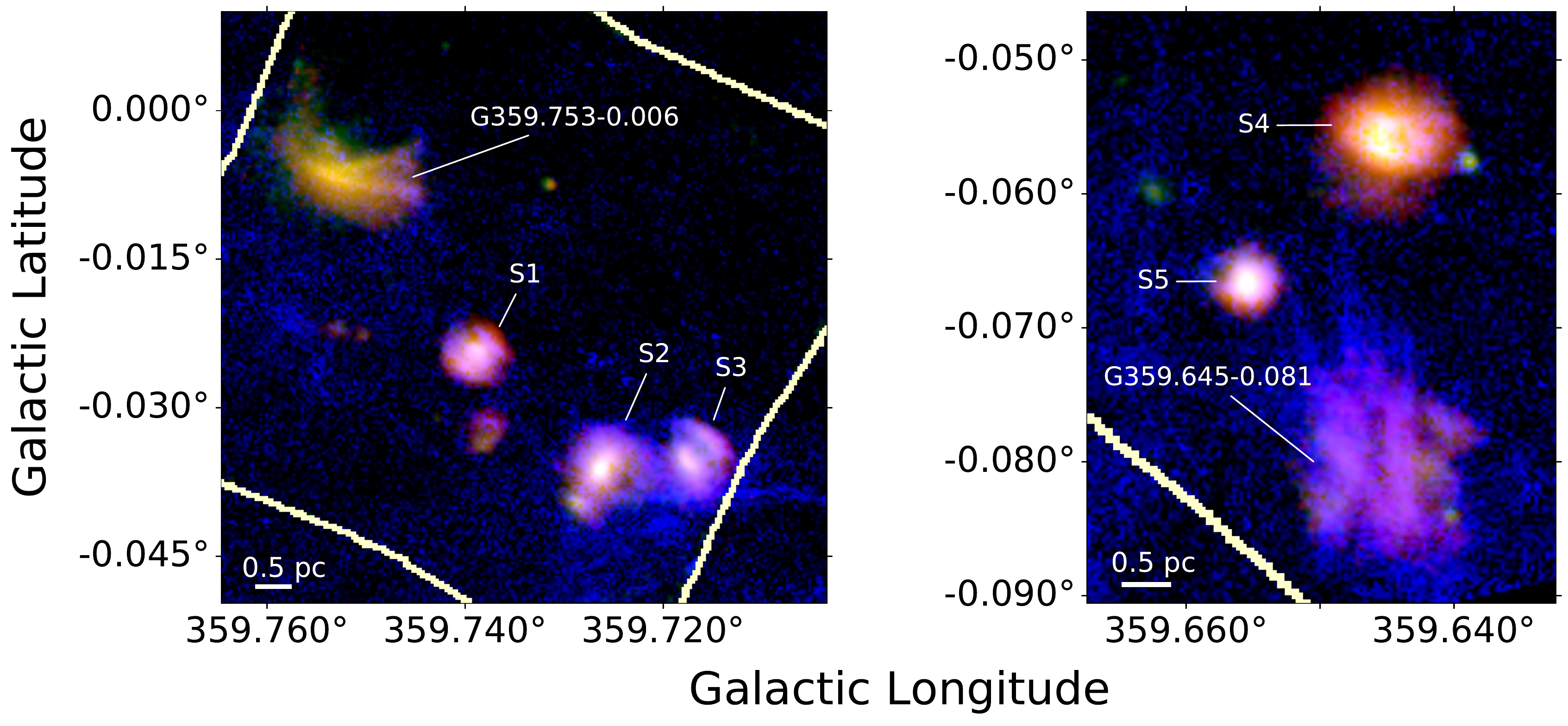}
\caption{{\footnotesize A collection of mid-IR sources located between Sgr A \& Sgr C. Both panels are false-color images created using \textit{HST} Paschen-$\alpha$ emission (blue), \textit{SOFIA}/FORCAST 25 $\mu$m (green), and 37 $\mu$m (red) data. A few regions of interest discussed in the text are labeled and the \textit{SOFIA}/FORCAST survey footprint is also shown for reference. These less well-studied sources are discussed in Section \ref{sec:SgrC}.}}
\label{fig:fig10}
\end{figure*}

Between Sgr A and Sgr C there is a smattering of compact mid-infrared sources which have not garnered as much attention as other portions of the GC. These sources appear to be in somewhat isolated environments, which speaks to the apparent dichotomy between regions located at negative and positive Galactic longitudes, the latter of which are thought to be more actively forming stars \citep{Longmore2013,Barnes2017}. These isolated western sources are of interest because several are cataloged as candidate YSOs from the ISOGAL survey \citep{Immer2012ISOgal}. Our observations of a handful of these sources show a diverse range of morphologies and colors. For example, G359.753-0.006 shows a prominent bow shock-like morphology (Figure \ref{fig:fig10}). Several other nearby sources (G359.738-0.024, G359.726-0.037, G359.716-0.035, G359.64477-0.056, and G359.655-0.067, labeled respectively as S1, S2, S3, S4, and S5 in Figure \ref{fig:fig10}) are more compact but still resolved in the FORCAST observations. These sources each show ionized gas emission and are possibly compact or ultra-compact \ion{H}{2} regions. Interestingly, S2 and S3 may be interacting with a nearby ionized gas filament which is visible in the Paschen-$\alpha$ data, and may have some similarities to the G359.866+0.002 complex discussed in section \ref{sec:SgrA}.

One of the most interesting sources in this region is G359.645-0.081, which is a relatively faint extended source. It is brighter at 37 $\mu$m compared to 25 $\mu$m, indicating that it is somewhat cool and is plausibly the edge of a molecular cloud. However, the amount of extended ionized gas emission in addition to the relatively cool dust emission is somewhat peculiar compared to other extended sources in our survey. Prior observations of this region with \textit{Spitzer}/IRS show a few high ionization species including [\ion{Ne}{5}] 24.32 $\mu$m and [\ion{O}{4}] 25.9 $\mu$m near this position \citep{Simpson2018IRS}, indicative of a hard ionization source. The morphology and excitation of G359.645-0.081 is somewhat reminiscent of the Sgr A East ejecta \citep{Lau2015}, and also to the supernova remnant G292.0+1.8 \citep{Ghavamian2009}; however, this region lacks a X-ray component \citep{Simpson2018IRS} suggesting it is not a supernova remnant. Although the unusual color and ionization properties of this source merit further study.

\section{Summary} \label{sec:Summary}

In this paper we have presented observations and initial results from the \textit{SOFIA}/FORCAST Survey of the GC. Our survey focused on some of the brightest infrared regions in the GC which trace recent star formation. These data provide the highest spatial resolution mapping of the Galactic Center at 25 and 37 $\mu$m to date (FWHM$\sim$2.3'' and FWHM$\sim$3.4'', respectively), and cover several interesting regions which were badly saturated in the \textit{Spitzer}/MIPS 24 $\mu$m data. Ultimately, our primary science objective for the survey is to better characterize star formation in the GC, in particular the well-known star formation rate discrepancy in this region. Examining this topic will require high-level data products, including source catalogs and enhanced maps that will be produced as a part of this program and released to the broader astronomical community in a future data release. 

We have produced this paper to describe the survey plan, observations, and initial highlights from the data as an accompaniment to the initial survey data release. All data collected from this survey, including level 3 and 4 data products, can be found on the SOFIA DCS and IRSA. In this work, we have presented short summaries of several featured regions and sources including extended structures near the CND, the Arched filaments and Sickle \ion{H}{2} regions, and embedded star formation in Sgr B2 and Sgr C. While primarily qualitative, these case studies illustrate the scope and utility of the survey data. We are planning future studies around several of these topics which will be presented in later works or possibly accompany future data releases.

\vspace{3mm}
\emph{Acknowledgments} We thank the anonymous referee for their comments which improved the quality of this paper. Additionally, we thank the USRA Science and Mission Ops teams and the entire SOFIA staff for making this survey possible. In particular, we thank Mike Gordon and Jim De Buizer who helped to support our observations. Additionally, we thank the many people who have worked on FORCAST over the years including but not limited to George Gull, Justin Schoenwald, Chuck Henderson, Joe Adams, and Andrew Helton. Financial support for this work was provided by NASA through award number NNA17BF53C issued by USRA. This material is based upon work supported by the National Science Foundation under grant No. AST-1813765. ATB would like to acknowledge the funding provided from the European Union's Horizon 2020 research and innovation programme (grant agreement No 726384).

This work is based on observations made with the NASA/DLR Stratospheric Observatory for Infrared Astronomy (SOFIA). SOFIA science mission operations are conducted jointly by the Universities Space Research Association, Inc. (USRA), under NASA contract NAS2-97001, and the Deutsches SOFIA Institut (DSI) under DLR contract 50 OK 0901. Financial support for FORCAST was provided by NASA through award 8500-98-014 issued by USRA.


This work made use of data products from the Spitzer Space Telescope, which is operated by the Jet Propulsion Laboratory, California Institute of Technology, under a contract with NASA. Additionally, this research made use of data products from the Midcourse Space Experiment. Processing of the data was funded by the Ballistic Missile Defense Organization with additional support from NASA Office of Space Science. This research has also made use of the NASA/ IPAC Infrared Science Archive, which is operated by the Jet Propulsion Laboratory, California Institute of Technology, under contract with the National Aeronautics and Space Administration. Finally, this research has also made use of the VizieR catalog access tool, CDS, Strasbourg, France. The original description of the VizieR service was published in A\&AS 143, 23.

\bibliographystyle{aasjournal}
\bibliography{main}

\end{document}